\documentclass[11pt,reqno]{amsart}

\usepackage{amsmath}
\usepackage{amsthm}
\usepackage{amssymb}
\usepackage{amsfonts}
\usepackage{fullpage}
\usepackage[latin1]{inputenc}

\usepackage{graphicx}

\def\be{\begin{equation}}
\def\ee{\end{equation}}
\def\bea{\begin{eqnarray}}
\def\eea{\end{eqnarray}}

\newcommand{\bq}{\mathbf{q}}
\newcommand{\bp}{\mathbf{p}}

\def\X{\mathbf{X}}
\def\Y{\mathbf{Y}}
\def\pr{{\rm pr}}
\def\Esp{{\mathcal E}}
\def\L{{\mathcal L}}

\def\dd{{\rm d}}
\def\sgn{{\rm sgn}}
\def\com/{c.o.m.}
\def\eom/{equations of motion}
\def\eff{{\rm eff}}

\begin{document}
\allowdisplaybreaks[3]

\title{}
\begin{center}
{\LARGE
\bf{Global versus local superintegrability \\ of nonlinear oscillators}
}
\end{center}
\bigskip

\begin{center}
{\large
{\sc Stephen C. Anco$^a$, Angel Ballesteros$^b$, Maria Luz Gandarias$^c$}
}
\bigskip
\def\1{\'{\i}}

{$^a$Department of Mathematics and Statistics, Brock University,
St Catharines, Canada\\ ~~E-mail: sanco@brocku.ca \\[10pt]}
{$^b$Departamento de F\1sica,  Universidad de Burgos,
09001 Burgos, Spain\\ ~~E-mail: angelb@ubu.es \\[10pt]}
{$^c$Departamento de Matem\'aticas, Universidad de C\'adiz,
Puerto Real, Spain \\ ~~E-mail: marialuz.gandarias@uca.es \\[10pt]}
\end{center}

\begin{abstract}
Liouville (super)integrability of a Hamiltonian system of differential equations
is based on the existence of globally well-defined constants of the motion, 
while Lie point symmetries provide a local approach to conserved integrals. 
Therefore, it seems natural to investigate in which sense Lie point symmetries 
can be used to provide information concerning the superintegrability of a given Hamiltonian system. 
The two-dimensional oscillator and the central force problem are used as benchmark examples 
to show that the relationship between standard Lie point symmetries and superintegrability 
is neither straightforward nor universal. 
In general, it turns out that superintegrability is not related to either the size or the structure of the algebra of variational dynamical symmetries.
Nevertheless, all of the first integrals for a given Hamiltonian system can be obtained 
through an extension of the standard point symmetry method, 
which is applied to a superintegrable nonlinear oscillator 
describing the motion of a particle on a space with non-constant curvature and spherical symmetry.\\


\end{abstract}

\maketitle

%%%%%%%%%%%%%%%%%%%%%%%%%%%%%%%%%%%%%%%%%%%%%%%%

\section{Introduction}\label{intro}

Hamiltonian systems of differential equations are of widespread importance in physics and mathematics. 
There is particular interest in systems that are superintegrable in the sense of Liouville
by possessing the maximal number of globally well-defined first integrals 
which are functionally independent. 

Noether's theorem provides an obvious connection between first integrals and local symmetries of a given Hamiltonian system, whether or not the system is (super)integrable. 
Specifically, using the Lagrangian formulation of the system, 
each first integral corresponds to a variational local symmetry, 
which can be either a point symmetry or a dynamical symmetry \cite{BA-book,Olver-book}. 
Point symmetries are distinguished by involving only the canonical variables and the time variable; 
most importantly, 
all point symmetries can be obtained systematically for any Hamiltonian system 
through the use of Lie's method \cite{BA-book,Olver-book}. 

A natural first question is: 
\emph{To what extent can (super)integrability of a Hamiltonian system be detected just by looking at its point symmetries?}
The answer is, in general, that variational point symmetries do not always provide 
a sufficient number of first integrals. 

A widely studied example is central force motion in the Newtonian case of an inverse-square force law (see e.g.~\cite{GolPooSaf}). 
The variational point symmetries of this Hamiltonian system consist of rotations and time-translation,
which yield the components of the angular momentum vector and the energy 
as first integrals. 
There are additional first integrals given by the components of the well-known Laplace-Runge-Lenz (LRL) vector. 
Recall that this vector lies in the plane orthogonal to the angular momentum vector 
and is oriented in the direction of the apsis line from the center of mass to the apsis point on any non-circular orbit. 
The first integrals corresponding to the angle determined by the LRL vector 
arise from ``hidden'' dynamical symmetries \cite{BacRueSou,Fra,AncMeaPas} 
rather than point symmetries of the \eom/. 

A second question then is: 
\emph{Can (super)integrability of a Hamiltonian system be detected by knowing its dynamical symmetries?}
The answer, surprisingly, is neither simple nor universal. 

Because variational dynamical symmetries correspond to first integrals by Noether's theorem,
these local symmetries do contain some information about the first integrals.
In some situations, the global form of the symmetry group transformations 
acting on solutions may indicate if the first integrals are globally single-valued and non-singular.
But, in general, this global question about first integrals is distinct from the properties of the variational symmetry group, 
because the existence and nature of local symmetries depends solely on the local structure of the \eom/.

The example of central force motion with a general radial potential is a nice illustration of the subtleties. 
It has been known for several decades that an analog of the LRL vector exists 
for any radial potential \cite{BacRueSou,Fra,AncMeaPas}, 
but the resulting first integrals given by the components of this generalized vector 
are globally single-valued and non-singular only for 
the Kepler-Coulomb potential and the isotropic oscillator potential \cite{BacRueSou}. 
Namely, those are the only two central force systems that are superintegrable 
in Euclidean space. 
(For the situation in curved spaces with radial symmetry, see~\cite{commun}.) 
When any other central force system is considered in Euclidean space, 
the generalized LRL vector instead is multi-valued and jumps each time the apsis point on a non-circular orbit is reached \cite{SerSha,BucDen}. 
Moreover, the ``hidden'' dynamical symmetries that correspond to these first integrals 
have the same symmetry algebra \cite{BacRueSou,Fra} for all radial potentials, 
whether the central force system is superintegrable or not. 
As a consequence, superintegrability is not related to either the size or the algebra structure of the variational dynamical symmetries. 

Of course, if all variational dynamical symmetries of a given Hamiltonian system are known,
then the first integrals can be obtained in an explicit form through Noether's theorem,
so that their global properties then can be studied. 
But a priori it is not possible to find all variational dynamical symmetries 
without in essence integrating the \eom/ of the Hamiltonian system,
and this task involves the same level of difficulty as directly finding all first integrals
\cite{BA-book}. 
More specifically, it is no easier to find all variational dynamical symmetries 
than it is to find all first integrals. 
Nevertheless, there is an extended symmetry method developed in Ref.~\cite{AncMeaPas} that can be used 
to obtain all first integrals systematically for many Hamiltonian systems.

The purpose of the present paper is to get a deeper insight into the connection between first integrals and local symmetries 
by studying a superintegrable Hamiltonian system introduced in Ref.~\cite{DarbouxPhysD}.
This system describes a radially symmetric nonlinear oscillator which is physically interesting both from dynamical and geometrical viewpoints, 
since it can be identified both 
with the motion of a particle on a space having non-constant curvature,
and also with an oscillator whose mass is position dependent. 

The main results in the paper will be to show how to go systematically 
from the Hamiltonian system to local symmetries and then to first integrals,
and reciprocally, from the Hamiltonian system directly to first integrals and then to local symmetries. 
This will be accomplished without the need for ansatzes or guess-work
by adapting the extended symmetry method from \cite{AncMeaPas}. 
As will be seen, 
superintegrability of the system is not related in any straightforward or universal way to
either its point symmetries or its variational symmetry algebra. 

These results will reinforce the preceding discussion. 
In particular,
on one hand, point symmetries are generally insufficient to characterize when a Hamiltonian system is superintegrable, 
and on the other hand, 
the size and structure of the variational symmetry algebra of a given Hamiltonian system 
is not enough to detect if the system is superintegrable.

%%%%%%%%%%%%%%%%%%%%%%%%%%%%%%%%%%%%%%%%%%%%%%%%

\section{Lie point symmetries and (super)integrability}\label{toymodels}

In this section,
the basic Hamiltonian systems for two uncoupled oscillators
and for central force planar motion 
are studied as benchmark models to discuss and clarify the questions raised in section~\ref{intro}.
Recall that the first system is superintegrable provided that the two oscillator frequencies are commensurate, 
and that the second system is superintegrable only for the Kepler-Coulomb potential and the isotropic oscillator potential. 
This situation is ideal for analyzing the difficulties in detecting superintegrability through point symmetries, 
since the superintegrable systems are specific cases belonging to the respective family of oscillator and central force potentials.

\subsection{Uncoupled oscillators}\label{uncoupledoscil}

The Hamiltonian system describing two uncoupled oscillators is given by 
the \eom/
\begin{equation*}
\ddot q_1 +\omega_1{}^2 q_1=0,
\qquad
\ddot q_2 +\omega_2{}^2 q_2=0,
\end{equation*}
for $(q_1(t),q_2(t))$,
where $\omega_1$ and $\omega_2$ are the frequencies.
The Lagrangian is $\L = \tfrac{1}{2}(\dot q_1^2 +\dot q_2^2 - \omega_1{}^2 q_1^2 - \omega_2{}^2 q_2^2)$. \\

\emph{First integrals}:
All constants of motion (\com/) $I(q_1,q_2,\dot q_1,\dot q_2)$ 
arise from the determining equation
\begin{equation*}
0=\dot I =I_{q_1} \dot q_1 +I_{q_2} \dot q_2 -\omega_1{}^2 q_1 I_{\dot q_1} -\omega_2{}^2 q_2 I_{\dot q_2},
\end{equation*}
which can be solved easily by the method of characteristics.
A maximal set of three functionally-independent \com/ is given by
the energies of the two oscillators
\be\label{uncoupledoscil-com-E}
E_1= \tfrac{1}{2}( \dot q_1^2 +\omega_1{}^2 q_1^2 ),
\quad
E_2= \tfrac{1}{2}( \dot q_2^2 +\omega_2{}^2 q_2^2 ) ,
\ee
and a phase quantity
\be\label{uncoupledoscil-com-Phi}
\Phi= (1+\tfrac{\omega_2}{\omega_1})\arctan(\omega_1 q_1/\dot q_1)  -(1+\tfrac{\omega_1}{\omega_2})\arctan(\omega_2 q_2/\dot q_2)
\ \mod \pi.
\ee
This quantity can be shown to describe
the difference in the relative phase shifts $\Delta\phi_1$ and $\Delta\phi_2$
between the two oscillators measured at the times $t_1$ and $t_2$
when each one of oscillators passes through zero,
namely $\Phi=\Delta\phi_2-\Delta\phi_1$.
It undergoes a jump each time one of the oscillators changes direction.
Thus, in general the \com/~$\Phi$ is multi-valued. 
In contrast, the \com/~$E_1$, $E_2$, and $E=E_1+E_2$ (total energy of the oscillators) are single-valued. 

The oscillator system is superintegrable iff the frequencies of two oscillators are commensurate:
$\omega_1/\omega_2\in\mathbb Q$.
In the superintegrable case,
the set of values of the \com/ $\Phi$ is finite, 
whereas for incommensurate frequencies,
this set of values is infinite. 
In both cases, $\Phi$ is always non-singular.

In addition to the three functionally-independent \com/ $E_1$, $E_2$, $\Phi$, 
there is a first integral that depends explicitly on $t$:
\be
T= t- \Big(\tfrac{1}{2\omega_1}\arctan(\omega_1 q_1/\dot q_1)  +\tfrac{1}{2\omega_2}\arctan(\omega_2 q_2/\dot q_2)\Big)
\ee
This first integral yields the average of the times at which the respective oscillators pass through zero. \\

\emph{Variational symmetries}:
According to Noether's theorem,
each of the first integrals $E_1$, $E_2$, $\Phi$, $T$ corresponds to a variational symmetry.
The evolutionary form of these symmetries, acting on $(q_1(t),q_2(t))$, is given by the generators 
\begin{gather*}
\hat\X_{E_1}= \dot q_1\partial/\partial q_1,
\quad
\hat\X_{E_2}= \dot q_2\partial/\partial q_2,
\\
\hat\X_{\Phi}=\tfrac{\omega_1+\omega_2}{2}( \tfrac{1}{E_2} q_2\partial/\partial q_2 -\tfrac{1}{E_1} q_1\partial/\partial q_1 ),
\quad
\hat\X_{T}=\tfrac{1}{2E_1} q_1\partial/\partial q_1 +\tfrac{1}{2E_2} q_2\partial/\partial q_2 .
\end{gather*}
By comparison, an infinitesimal point symmetry 
$t\to t + \epsilon \tau(t,q_1,q_2) +O(\epsilon^2)$,
$q_1\to q_1+\epsilon \eta_1(t,q_1,q_2) +O(\epsilon^2)$,
$q_2\to q_2+\epsilon \eta_2(t,q_1,q_2) +O(\epsilon^2)$
in evolutionary form has the generator
%$\X = \tau(t,q_1,q_2)\partial/\partial_t + \eta_1(t,q_1,q_2)\partial/\partial_{q_1}+ \eta_2(t,q_1,q_2)\partial/\partial_{q_2}$ 
$\hat\X=(\eta_1-\tau\dot q_1)\partial/\partial_{q_1}+ (\eta_2-\tau\dot q_2)\partial/\partial_{q_2}$.
It is straightforward to see that none of the four symmetries $\hat\X_{E_1}$, $\hat\X_{E_2}$, $\hat\X_{\Phi}$, $\hat\X_{T}$
represent point symmetries, due to the form of their dependence on $\dot q_1$ and $\dot q_2$ through the expressions \eqref{uncoupledoscil-com-E}--\eqref{uncoupledoscil-com-Phi} for $E_1,E_2,\Phi$.
Therefore, they are dynamical symmetries.

On solutions of the \eom/,
the symmetry generators are mutually commuting,
namely the symmetry algebra is abelian. 
Moreover, none of the symmetries contain information about superintegrability of the \eom/. 
In particular, the components of each symmetry generator are
single-valued and non-singular for arbitrary frequencies $\omega_1\neq0$, $\omega_2\neq0$.\\

\emph{Point symmetries}:
The point symmetries of the uncoupled oscillator system
for general frequencies $\omega_1$ and $\omega_2$
are generated by
a time translation $\X_{\rm trans}= \partial/\partial t$
and two scalings $\X_{{\rm scal}_1}= q_1\partial/\partial q_1$ and $\X_{{\rm scal}_2}= q_2\partial/\partial q_2$, 
along with 
elementary symmetries $\X = f_1(t)\partial/\partial_{q_1} + f_2(t)\partial/\partial_{q_2}$,
where $f_j=a_j\cos(\omega_jt +\phi_j)$, $j=1,2$, are arbitrary solutions of the \eom/.
Note that time translation is a variational symmetry whose evolutionary form
is given by $\hat\X_{\rm trans} =\hat\X_{E_1} +\hat\X_{E_2} = \hat\X_{E}$. 

Additional point symmetries arise only when the two frequencies are equal,
$\omega_1=\omega_2=\omega$.
In this special case, 
there are four additional point symmetries,
which are generated by 
\begin{align*}
& \X_{\rm rot}= q_2\partial/\partial q_1 - q_1\partial/\partial q_2,
\\
& \X_1 = e^{\pm i\omega t}( q_1\partial/\partial q_1 + q_2\partial/\partial q_2 \mp\tfrac{i}{\omega} \partial/\partial_t ),
\\
& \X_2 = e^{\pm i\omega t}q_1(q_1\partial/\partial q_1 + q_2\partial/\partial q_2 \mp\tfrac{i}{\omega} \partial/\partial_t ),
\\
& \X_3 = e^{\pm i\omega t}q_2(q_1\partial/\partial q_1 + q_2\partial/\partial q_2 \mp\tfrac{i}{\omega} \partial/\partial_t ).
\end{align*}

Hence, because $\omega_1=\omega_2=\omega$ belongs to the case of commensurate frequencies,
this special superintegrable case has a larger point symmetry group.
However, in all other superintegrable cases,
for which $\omega_1/\omega_2\in \mathbb Q$ with $\omega_1\neq\omega_2$,
the point symmetry group of the system has the same size as in the general non-superintegrable case.

%%%%%%%%%%%%%%%%%%%%%%%%%%%%%%%%%%%%%%%%%%%%%%

\subsection{Central force motion}\label{centralforce}

For any central force, 
motion in the plane orthogonal to the conserved angular momentum vector
is given by the Hamiltonian system 
\begin{equation*}
\ddot r=\dot \theta^2 r -U'(r), 
\quad
\ddot \theta=-2\dot \theta \dot r/r
\end{equation*}
in polar coordinates $(r(t),\theta(t))$, 
where $U(r)$ is the potential and $\L = \tfrac{1}{2}( \dot r^2  + \dot\theta^2 r^2 ) -U(r)$ is the Lagrangian.\\

\emph{First integrals}:
All \com/ $I(r,\theta,\dot r,\dot \theta)$ arise from the determining equation
\begin{equation*}
0=\dot I =I_{r} \dot r +I_{\theta} \dot \theta +(\dot\theta^2 r -U'(r)) I_{\dot r} -2\dot\theta r^{-1}\dot r I_{\dot \theta} .
\end{equation*}
This is a first-order linear PDE for $I$, which can be explicitly solved by the method of characteristics
and yields a maximal set of three functionally-independent \com/:
\begin{gather}
L=\dot\theta r^2, 
\quad
E= \tfrac{1}{2}( \dot r^2 +L^2/r^2) +U(r), 
\label{centralforce-LE}
\\
\Theta= \theta-L\int_{r_0}^r \frac{\sgn(\dot r)}{r\sqrt{2(E+U(r_{\rm equil}) - U(r))r^2 -L^2}}dr
\ \mod 2\pi,
\label{centralforce-Theta}
\end{gather}
where $r_{\rm equil}$ is any equilibrium point, $U'(r_{\rm equil})=0$. 
For any solution $(r(t),\theta(t))$ of the \eom/, 
$L$ is the planar angular momentum;
$E$ is the energy (Hamiltonian);
and $\Theta$ is the angle reached at some point $r=r_0$.
As shown in \cite{AncMeaPas},
a natural intrinsic choice of $r_0$ is any turning point $r^*$ or any inertial point $r_*$, 
which are given by $U_\eff(r^*)=E$ or $U_\eff'(r_*)=0$
in terms of the effective potential $U_\eff(r)=U(r) +\tfrac{1}{2}L^2/r^2 - U(r_{\rm equil})$.

In addition to the three functionally-independent \com/~ $L$, $E$, $\Theta$, 
there is a first integral that depends explicitly on $t$:
\be\label{centralforce-T}
T = t- \int^r_{r_0} \frac{\sgn(\dot r)}{\sqrt{2(E+U(r_{\rm equil})-U(r)) -L^2/r^2}}\,dr .
\ee

It is well known that the central force system is superintegrable iff
$U(r)=-k/r$ is the Coulomb potential or $U(r)=k r^2$ is the isotropic oscillator potential.
Thus, $\Theta$ is single-valued and non-singular in these two cases.
In particular, if $r_0=r^*$ is a turning point at which $r(t)$ reaches a local maximum,
then as shown in \cite{AncMeaPas},
$\Theta$ is the angle of the LRL vector,
which is a \com/ for the Coulomb potential and the isotropic oscillator potential. 
In these two cases, $T$ is the time at which $\theta(t)$ coincides with the LRL angle,
modulo the period of the solution $(r(t),\theta(t))$.
As a consequence, all bounded orbits for both of these potentials do not precess. 
For any other central force system, we can infer that $\Theta$ is multi-valued and possibly singular.
In particular, the angle $\Theta$ defining the generalized LRL vector undergoes a jump each time
$t=T$ when $r$ reaches a turning point.
An example is the perturbed Coulomb potential $U(r)=-k/r-K/r^2$,
where bounded non-circular orbits exhibit precession \cite{GolPooSaf}, 
and thus this central force system is not superintegrable. \\

\emph{Variational symmetries}:
By Noether's theorem,
each of the \com/~$L$, $E$, $\Theta$ corresponds to a variational symmetry.
In evolutionary form, acting on $(r(t),\theta(t))$,
these symmetries are given by the generators \cite{AncMeaPas}
\begin{gather*}
\hat\X_{L}= -\partial/\partial \theta,
\quad
\hat\X_{E}= -\dot r\partial/\partial r -\dot\theta \partial/\partial \theta,
\\
\hat\X_{\Theta}=-(\dot r \partial_E\Theta) \partial/\partial r -(\partial_L\Theta +\dot\theta\partial_E \Theta)\partial/\partial \theta,
\quad
\hat\X_{T}=-(\dot r \partial_E T) \partial/\partial r -(\partial_L T +\dot\theta\partial_E  T)\partial/\partial \theta .
\end{gather*}
Both $\hat\X_{L}$ and $\hat\X_{E}$ represent infinitesimal point symmetries,
as can be easily seen by comparison with
$\hat\X=(\eta^r -\tau\dot r)\partial/\partial{r}+ (\eta^\theta-\tau\dot \theta)\partial/\partial{\theta}$
which is general evolutionary form for an infinitesimal point symmetry 
$t\to t + \epsilon \tau(t,r,\theta) +O(\epsilon^2)$,
$r\to r+\epsilon \eta^r(t,r,\theta)  +O(\epsilon^2)$,
$\theta\to \theta+\epsilon \eta^\theta(t,r,\theta)  +O(\epsilon^2)$. 
In contrast, $\hat\X_{\Phi}$ and $\hat\X_{T}$ do not represent point symmetries
but instead are dynamical symmetries,
because of their nonlinear dependence on $\dot r$ and $\dot \theta$ through the expressions \eqref{centralforce-LE} for $L,E$.

A straightforward computation shows that, on solutions of the \eom/,
the symmetry algebra is abelian,
namely, the four generators $\hat\X_{L}$, $\hat\X_{E}$, $\hat\X_{\Phi}$, $\hat\X_{T}$
are mutually commuting.

Clearly, the point symmetries $\X_{L}$ and $\X_{E}$ contain no information about superintegrability of the \eom/,
since they are admitted for an arbitrary central force potential $U(r)$.
An interesting question is whether the dynamical symmetry $\X_{\Theta}$ contains
any information about superintegrability.
To answer this, we need to examine the components
\be\label{centralforce_Theta_components}
\begin{aligned}
\partial_E \Theta & = 
L\int_{r_0}^r \frac{\sgn(\dot r)r}{\sqrt{2(E+U(r_{\rm equil}) - U(r))r^2 -L^2}^3}dr,
\\
\partial_L\Theta & =
\int_{r_0}^r \frac{\sgn(\dot r)2(U(r)-U(r_{\rm equil})-E)r}{\sqrt{2(E+U(r_{\rm equil}) - U(r))r^2 -L^2}^3}dr.
\end{aligned}
\ee
In the cases of the Coulomb potential $U(r)=-k/r$
and the isotropic oscillator potential $U(r)=k r^2$,
we find that both components \eqref{centralforce_Theta_components}
are single-valued but become singular at turning points.
For the case of the perturbed Coulomb potential $U(r)=-k/r-K/r^2$,
we find that the component $\partial_L\Theta$ is not single-valued. 
Consequently, in these three examples,
the form of the dynamical symmetry $\X_{\Theta}$ detects if the central force system is superintegrable. \\

\emph{Point symmetries}:
The point symmetries of the central force \eom/ are generated by 
$\X_{L}= -\partial/\partial \theta$ and $\X_{E}= \partial/\partial t$
for a general potential $U(r)$. 
Additional point symmetries are admitted only for \cite{AncMeaPas}
two special potentials:
$U(r)=kr^p$, which admits $\X_1 = t\partial/\partial t -\tfrac{2}{p}r\partial/\partial r$;
$U(r)=kr+K/r^3$, which admits $\X_2 = e^{2\sqrt{k}t}(\partial/\partial t +\sqrt{k}r\partial/\partial r)$.

Notice that the point symmetry group is \emph{not larger} in the superintegrable cases.

%%%%%%%%%%%%%%%%%%%%%%%%%%%%%%%%%%%%%%%%%%%%%%%%%%%%%%%%%%%%%%%%%%%%%%%%%%%%

\section{Connections among first integrals, symmetries, and superintegrability}

We will study the $N=2$ version of the dynamical system given by the Hamiltonian 
\be\label{ND-Ham}
H(\bq,\bp)
%=\frac{\bp^2}{2(1+\la \bq^2)} + \frac{ \omega^2 \bq^2}{2(1+\la \bq^2)}
=\tfrac{1}{2}(1+\lambda \bq^2)^{-1} ( \bp^2 + \omega^2 \bq^2)
\ee 
where $\lambda>0$ and $\omega> 0$ are real parameters,
and $(\bq,\bp)$ are $2N$ canonical coordinates. 
This system  was proven in~Ref.~\cite{DarbouxPhysD} to be maximally superintegrable,
namely it possesses the maximum number $(2N-1)$ of \com/,
which are functionally independent and globally well-defined for a general solution $(\bq(t),\bp(t))$.
These \com/ are explicitly given by 
\be\label{ND-com}
\begin{gathered}
C^{(m)}=\!\! \sum_{1\leq i<j\leq m} \!\!\!\! (q_ip_j-q_jp_i)^2 ,
\quad 
C_{(m)}=\!\!\! \sum_{N-m<i<j\leq N}\!\!\!\!\!\!  (q_ip_j-q_jp_i)^2 , \quad m=2,\dots,N,
\\
E_i=p_i^2-\bigl(2\lambda  H(\bq,\bp)-\omega^2\bigr) q_i^2 ,\quad i=1,\dots,N.
\end{gathered}
\ee
Some non-local symmetries for the $N=1$ case were found in~\cite{GandariasPLA}.

When $\lambda\to 0$, 
this system~\eqref{ND-Ham} reduces to the $N$-dimensional Euclidean isotropic oscillator with frequency $\omega$, 
which is indeed a maximally superintegrable system.
Thus, $\lambda$ can be viewed as a deformation parameter, 
and the system~\eqref{ND-Ham} can be thought of as a maximally superintegrable deformation of the Euclidean isotropic oscillator.
Geometrically, 
the term $\tfrac{1}{2}(1+\lambda \bq^2)^{-1}\bp^2$ in the Hamiltonian $H(\bq,\bp)$
can be interpreted as the kinetic energy defined by the geodesic motion of a particle with unit mass on a conformally flat space whose metric is given by 
$\dd s^2= (1+\lambda \bq^2)\dd \bq^2$ (see also~\cite{KKMW02}--\cite{Gonera}).  
The scalar curvature of this space
is negative and asymptotically vanishes for large $|\bq|$.
Further discussion on the geometrical interpretation of the system and its quantization can be found in~\cite{commun} and~\cite{NDAnnals}--\cite{BGSNpla}.
From a physical viewpoint,
the system describes a particle with position-dependent mass of the form $m(\bq)=1+\lambda \bq^2$. 
We recall that the quantum version of such systems (see for instance~\cite{Roos}--\cite{MR}  and references therein) is relevant for the description of
semiconductor heterostructures and nanostructures
and, in particular, models constructed in terms of quadratic mass functions have been considered in~\cite{Koc, Schd}.

Our main result will be to show systematically how to derive the local symmetry group underlying the \com/ ~\eqref{ND-com} of this system in the case $N=2$. 
We do this derivation in two different ways.

First, we directly integrate the determining equation for first integrals, 
and then we apply Noether's theorem (in reverse) to obtain the corresponding variational symmetries.
This process can be summarized as
\emph{Hamiltonian system $\Rightarrow$ first integrals $\Rightarrow$ local symmetries}.

Next, and most importantly, 
we show how to use the symmetry method outlined in Ref.\cite{AncMeaPas}
to do the reverse process:
\emph{Hamiltonian system $\Rightarrow$ local symmetries $\Rightarrow$ first integrals}.
This method is systematic and explicit, and no ansatzes are needed.

\subsection{From symmetries to first integrals}

The Hamiltonian \eqref{ND-Ham} in the planar case $N=2$ is given by
\be\label{Ham}
H = \frac{p_1^2 +p_2^2 +\omega^2(q_1^2+q_2^2)}{2(1+\lambda(q_1^2+q_2^2))}.
\ee
Note, when the deformation parameter $\lambda$ is taken to be $\lambda=0$,
this Hamiltonian reduces to the one for the planar isotropic oscillator,
which is given by oscillator system discussed in Section~\ref{uncoupledoscil}
in the case $\omega_1=\omega_2=\omega$. 

Hereafter it will be useful to change from planar coordinates to polar coordinates 
\be 
q_1=r\sin\theta,
\quad
q_2=r\sin\theta,
\quad
p_1 = p^r\cos\theta +p^\theta r^{-1}\sin\theta,
\quad
p_2 = p^r\sin\theta-p^\theta r^{-1}\cos\theta.
\ee
The Hamiltonian \eqref{Ham} becomes 
\be
H= \frac{(p^r)^2 + (p^\theta/r)^2  +\omega^2 r^2}{2(1+\lambda r^2)} , 
\ee
which yields the second-order ODE system
\be\label{eom}
\ddot r = f^r(r,\theta,\dot r,\dot\theta)
= \frac{((2\lambda r^2 +1)\dot \theta^2 +\lambda \dot r^2)r}{\lambda r^2+1} -\frac{\omega^2 r}{(\lambda r^2+1)^3},
\quad
\ddot\theta =  f^\theta(r,\theta,\dot r,\dot\theta) 
=-\frac{2\dot \theta \dot r (2\lambda r^2 +1)}{(\lambda r^2 +1)r}.
\ee
This system is superintegrable.
Hereafter,
the set of solutions $(r(t),\theta(t))$ of the \eom/ will be denoted $\Esp$.

%%%%%%%%%%%%%%%%%%%%%%%%%%%%%%%%%%%%%%%%%%%%%%%%

\subsection{From first integrals to symmetries}

A first integral is a function $I$ of $t,r,\theta,\dot r,\dot \theta$ that is time-independent, 
$\dot I=0$,
when it is evaluated on the solution space $\Esp$ of the \eom/.
If $I$ does not depend explicitly on $t$, then it is a \com/. 

All first integrals can be found by solving the determining equation
\be\label{I-deteqn}
0=\dot I(t,r,\theta,\dot r,\dot \theta)\big|_\Esp
=I_{t} + I_{r} \dot r +I_{\theta} \dot\theta +f^r I_{\dot r} +f^\theta I_{\dot \theta},
\ee
which is a linear first-order PDE for $I(t,r,\theta,\dot r,\dot \theta)$.
Solving this PDE amounts to integrating the \eom/ \eqref{eom}.
This can be done by applying the method of characteristics,
yielding the ODE system
$dt/1=dr/\dot r=d\theta/\dot\theta=d\dot r/f^r=d\dot\theta/f^\theta$.
Integration of the system gives four functionally-independent first integrals:
\begin{align}
& L= r^2(1+\lambda r^2) \dot\theta =p^\theta,
\label{L}
\\
&
E= \tfrac{1}{2}\big( (1+\lambda r^2) \dot r^2 + (\omega^2 r^2 +L^2/r^2)(1+\lambda r^2) ^{-1} \big) =H,
\label{E}
\\
& \begin{aligned}
\Theta & = \theta - \tfrac{1}{2}\arctan\Big(\frac{\sgn(\dot r)(Er^2-L^2)}{L\sqrt{2Er^2(1+\lambda r^2)-L^2-\omega^2r^4}}\Big)\Big|^{r}_{r_0}
%+ \tfrac{1}{2}\arctan\Big(\frac{\sgn(\dot r)_0(Er_0{}^2-L^2)}{L\sqrt{2Er_0{}^2(1+\lambda r_0{}^2)-L^2-\omega^2r_0{}^4}}\Big)\
\mod 2\pi,
\end{aligned}
\label{Theta}
\\
& 
\begin{aligned}
T & = t -\tfrac{1}{2}\frac{w^2-\lambda E}{\sqrt{w^2-2\lambda E}^3}\bigg(\arctan\Big(\frac{\sgn(\dot r)\big( (2\lambda r^2+1)E-\omega^2 r^2\big)}{\sqrt{w^2-2\lambda E}\sqrt{2Er^2(1+\lambda r^2)-L^2-\omega^2r^4}}\Big) \Big|^{r}_{r_0} \bigg) 
%-\arctan\Big(\frac{\sgn(\dot r)_0\big( (2\lambda r_0{}^2+1)E -\omega^2 r_0{}^2 )}{\sqrt{w^2-2\lambda E}\sqrt{2Er_0{}^2(1+\lambda r_0{}^2)-L^2-\omega^2r_0{}^4}}\Big)\bigg)
\\&\qquad
+\tfrac{1}{2}\frac{\lambda}{w^2-2\lambda E}\Big(
\sgn(\dot r)\sqrt{2Er^2(1+\lambda r^2)-L^2-\omega^2r^4}\Big)\Big|^{r}_{r_0} .
%- \sgn(\dot r)_0\lambda\sqrt{2Er_0{}^2(1+\lambda r_0{}^2)-L^2-\omega^2r_0{}^4}
\end{aligned}
\label{T}
\end{align}
Here $L$ is the planar angular momentum
and $E$ is the energy, 
which respectively arise from solving $dr/\dot r=d\dot\theta/f^\theta$
and $dr/\dot r=d\dot r/f^r$; 
$\Theta$ is an angular quantity
given by solving $dr/\dot r=d\theta/\dot\theta$
and involves an arbitrary constant of integration $r_0=r(t_0)$.
These quantities are \com/,
while $T$ is a temporal first integral arising from $dt/1=dr/\dot r$.
The physical meaning of $\Theta$ and $T$ is similar to the analogous quantities that appear in the superintegrable cases of central force motion discussed in Section~\ref{centralforce}.
Notice that $L$, $E$, $\Theta$, $T$ are functionally independent because
they each have different physical units. 

A natural physical choice of $r_0$ is any turning point $r^*$ or any inertial point $r_*$,
which are respectively given by $U_\eff(r^*)=E$ or $U_\eff'(r_*)=0$
in terms of the effective potential
\be
U_\eff(r)=\frac{\omega^2 r^2 +L^2/r^2}{1+\lambda r^2}.
\ee
On the orbit of a solution $(r(t),\theta(t))$,
a turning point is thus a point $r=r^*$ at which the radial velocity $\dot r=0$,
and an inertial point is a point $r=r_*$ at which the radial acceleration $\ddot r=0$.
These points are determined intrinsically by the dynamics of each solution $(r(t),\theta(t))$.
With such a choice of $r_0$,
the angular quantity $\Theta$ physically represents the angle $\theta$
on the orbit of a solution $(r(t),\theta(t))$
at the point $r=r_0$
given by either a turning point $r_0=r^*$ or an inertial point $r_0=r_*$.
Likewise,
the temporal quantity $T$ physically represents the time $t$
at which this point is reached on the orbit.

Superintegrability of the system \eqref{eom} is distinguished by the feature that
both $\Theta$ and $T$ are single-valued and non-singular for a general solution $(r(t),\theta(t))$.
Consequently,
$\Theta$ provides an analog of the LRL angle. 

All first integrals $I(t,r,\theta,\dot r,\dot\theta)$ are associated to multiplier pairs
$(Q^r(r,\theta,\dot r,\dot\theta),Q^\theta(r,\theta,\dot r,\dot\theta))$
given by expressing the conservation property $\dot I|_\Esp=0$ as an identity 
\begin{equation*}
\dot I = (\ddot r - f^r)Q^r + (\ddot \theta - f^\theta)Q^\theta,
\qquad
Q^r=\partial_{\dot r}I, 
\quad
Q^\theta=\partial_{\dot\theta}I 
\end{equation*}
holding off of the solution space $\Esp$.
Multiplier pairs are directly related to variational symmetries through Noether's theorem
using the Lagrangian formulation of the system \eqref{eom} as follows.

The Lagrangian is given by
$\L = \tfrac{1}{2}(1+\lambda r^2)( \dot r^2  + \dot\theta^2 r^2 ) -\tfrac{1}{2}w^2r^2/(1+\lambda r^2)$,
which yields 
\begin{equation*}
\ddot r -f^r =-(1+\lambda r^2)^{-1}\frac{\delta\L}{\delta r},
\qquad
\ddot \theta -f^\theta =-(r^2 (1+\lambda r^2))^{-1}\frac{\delta\L}{\delta \theta} . 
\end{equation*}
Now consider any vector field
$\hat\X = P^r(t,r,\theta,\dot r,\dot\theta)\partial_r + P^\theta(t,r,\theta,\dot r,\dot\theta)\partial_\theta$
in evolutionary form which acts only on the coordinates $(r,\theta)$. 
This vector field induces a variation of the Lagrangian, yielding Noether's identity
\begin{equation*}
\pr^{(1)}\hat\X (\L) = 
\frac{\delta\L}{\delta r} P^r + \frac{\delta\L}{\delta\theta} P^\theta + \frac{d}{dt}( P^r\partial_{\dot r}\L  + P^\theta \partial_{\dot\theta}\L),
\end{equation*}
where 
$\pr^{(1)}\hat\X = P^r\partial_{r} + P^\theta\partial_{\theta} + \dot P^r\partial_{\dot r} + \dot P^\theta\partial_{\dot\theta}$ 
is the prolongation of $\hat\X$ to the coordinate space
$(r,\theta,\dot r,\dot\theta)$. 
The condition for the vector field to be a variational symmetry is that
the induced variation of the Lagrangian is a total time derivative, 
$\pr^{(1)}\hat\X(\L) = \dot R$,
for some function $R(t,r,\theta,\dot r,\dot\theta)$. 
This implies 
\begin{equation*}
(\ddot r -f^r)\big( (1+\lambda r^2)P^r \big)
+ (\ddot \theta -f^\theta)\big( r^2(1+\lambda r^2)P^\theta \big)
= \frac{d}{dt}\big( P^r\partial_{\dot r}\L  +P^\theta \partial_{\dot\theta}\L -R \big) .
\end{equation*}
When this equation is evaluated on solutions $(r(t),\theta(t))$ of the \eom/,
it yields a first integral
\be\label{I}
I=R - P^r\partial_{\dot r}\L  -P^\theta \partial_{\dot\theta}\L  
\ee
which has the multiplier pair 
\be\label{QfromP}
Q^r = -(1+\lambda r^2)P^r ,
\qquad
Q^\theta = -r^2(1+\lambda r^2)P^\theta .
\ee
This relation establishes a one-to-one correspondence between
multiplier pairs $(Q^r,Q^\theta)$ and components $P^r$, $P^\theta$ of variation symmetries. 
Specifically, 
any variational symmetry yields a first integral whose corresponding multiplier 
is determined by equation \eqref{QfromP} in terms of the components of the symmetry generator;
conversely, any first integral yields a variational symmetry whose generator has
components determined in terms of the multiplier through inverting equation \eqref{QfromP}
to get
\be\label{PfromQ}
P^r = \frac{-Q^r}{1+\lambda r^2}, 
\qquad
P^\theta = \frac{-Q^\theta}{r^2(1+\lambda r^2)} . 
\ee
These correspondences \eqref{QfromP}--\eqref{PfromQ},
along with the first integral expression \eqref{I}, 
constitute the statement of Noether's theorem. 

The multiplier pairs for the four first integrals \eqref{L}--\eqref{T} are given by 
\begin{align}
& (Q^r,Q^\theta)_L = \big(0,r^2(\lambda r^2+1)\big), 
\qquad
(Q^r,Q^\theta)_E = \big((\lambda r^2+1)\dot r,r^2(\lambda r^2+1)\dot \theta\big), 
\label{Q-L-E}
\\
& (Q^r,Q^\theta)_\Theta = \big((\lambda r^2+1)\dot r\partial_E\Theta,r^2(\lambda r^2+1)(\dot \theta \partial_E\Theta+ \partial_L\Theta)\big),
\label{Q-Theta}
\\
& (Q^r,Q^\theta)_T = \big((\lambda r^2+1)\dot r\partial_E T,r^2(\lambda r^2+1)(\dot \theta \partial_E T+ \partial_L T)\big),
\label{Q-T}
\end{align}
where
\begin{align}
&\begin{aligned} 
\partial_L\Theta & =
\frac{1}{E^2+(2E\lambda-\omega^2)L^2} \bigg( 
\sgn(\dot r)\frac{(2(\lambda r^2 +1)E -\omega^2 r^2)E+(2\lambda E-\omega^2)L^2 }{2\sqrt{2Er^2(\lambda r^2 +1) - L^2 - \omega^2r^4}} \bigg)\bigg|^{r}_{r_0},
\end{aligned}
\\
&\begin{aligned}
\partial_E\Theta = - \partial_L T &= 
\frac{L}{E^2+(2E\lambda-\omega^2)L^2} \bigg( 
\sgn(\dot r)\frac{\omega^2r^2 -(\lambda r^2 +1)E -\lambda L^2}{2\sqrt{2Er^2(\lambda r^2 +1) - L^2 - \omega^2r^4}} \bigg)\bigg|^{r}_{r_0},
\end{aligned}
\\
&\begin{aligned}
\partial_ET & =\frac{\lambda(2w^2-\lambda E)}{2\sqrt{w^2-2\lambda E}^5}
\arctan\Big(\frac{\sgn(\dot r)\big((2\lambda r^2+1)E -\omega^2 r^2\big)}{\sqrt{2Er^2(\lambda r^2 +1)-L^2-\omega^2r^4}\sqrt{w^2-2\lambda E}}\Big)\Big|^{r}_{r_0}
\\&\qquad
+\frac{1}{2(w^2-2\lambda E)}
\Big( (\lambda r^2+1)\Big( \lambda r^2 +\frac{(\omega^2-\lambda E)(Er^2-L^2)}{E^2+(2E\lambda-\omega^2)L^2}\Big)\Big)\Big|^{r}_{r_0},
\\&\qquad
+\frac{\lambda}{2(w^2-2\lambda E)^2}
\Big(2\lambda +\frac{E(\omega^2-\lambda E)}{E^2+(2E\lambda-\omega^2)L^2}\Big)\Big(\sgn(\dot r)\sqrt{2Er^2(\lambda r^2 +1)-L^2-\omega^2r^4}\Big)\Big|^{r}_{r_0}.
\end{aligned}
\end{align}

Applying the Noether correspondence \eqref{PfromQ} to each multiplier pair, 
we obtain the corresponding variational symmetries
\begin{gather}
\hat\X_{L}= -\partial/\partial \theta,
\quad
\hat\X_{E}= -\dot r\partial/\partial r -\dot\theta \partial/\partial \theta,
\label{hatX-L-E}
\\
\hat\X_{\Theta} =-(\dot r \partial_E\Theta) \partial/\partial r -(\partial_L\Theta +\dot\theta\partial_E \Theta)\partial/\partial \theta,
\label{hatX-Theta}
\\
\hat\X_{T} =-(\dot r \partial_E T) \partial/\partial r -(\partial_L T +\dot\theta\partial_E  T)\partial/\partial \theta.
\label{hatX-T}
\end{gather}
These symmetries can be understood as acting on solutions $(r(t),\theta(t))$ of the \eom/.
Equivalently, the symmetries can be formulated as acting on the variables $(t,r,\theta)$
by means of a standard transformation \cite{BA-book,Olver-book}
which has the general form
$\eta^r=P^r +\tau \dot r$, $\eta^\theta=P^\theta +\tau \dot\theta$,
yielding 
\begin{equation*}
\X = \tau\partial/\partial_t + \eta^r\partial/\partial_r + \eta^\theta\partial/\partial_\theta
\end{equation*}
where $\tau$ can be chosen freely as a function of $t,r,\theta,\dot r,\dot\theta$. 
We take $\tau=0$ for $\hat\X_{L}$, $\tau=1$ for $\hat\X_{E}$,
giving
\be\label{X-L-E}
\X_{L}= -\partial/\partial \theta,
\qquad
\X_{E}= \partial/\partial t .
\ee
In this form, these generators represent point symmetries,
consisting of rotations and time-translations. 
For $\hat\X_{\Theta}$ we take $\tau = \partial_E\Theta$,
which yields
\be\label{X-Theta}
\begin{aligned}
\X_{\Theta} & =
\partial_E\Theta\partial/\partial_t -\partial_L\Theta\partial/\partial \theta .
\end{aligned}
\ee
Similarly, for $\hat\X_{T}$ we take $\tau = \partial_E T$,
giving 
\be\label{X-T}
\begin{aligned}
\X_{T} & =
\partial_E T \partial/\partial_t -\partial_L T\partial/\partial \theta .
\end{aligned}
\ee
Both of these generators $\X_{\Theta}$ and $\X_{T}$ represent dynamical symmetries.
Moreover, there is no choice of $\tau$ that can transform them into point symmetries,
because the components of the generators $\hat\X_{L}$ and $\hat\X_{T}$
have a nonlinear dependence on $\dot r$ and $\dot\theta$ through the expressions for $L,E$. 

The commutators of the variational symmetries \eqref{X-L-E}, \eqref{X-Theta}, \eqref{X-T}
turn out to vanish, as shown later, 
whereby the variational symmetries comprise a four-dimensional abelian algebra. 

In summary, 
we have shown how to go 
\emph{Hamiltonian system $\Rightarrow$ first integrals $\Rightarrow$ local symmetries}
in a systematic way by using the determining equation \eqref{I-deteqn} for first integrals
and the reverse version \eqref{PfromQ} of Noether's theorem.

\subsection{From symmetries to first integrals}

We will now show how to go 
\emph{Hamiltonian system $\Rightarrow$ local symmetries $\Rightarrow$ first integrals}
in a systematic way without having to use any ansatzes or guess-work,
by following the extended symmetry method outlined in Ref.~\cite{AncMeaPas}.

\emph{Step 1}:
Compute all variational point symmetries of the Hamiltonian system \eqref{eom}. 

Point symmetries are given by generators (vector fields) of the form  
$\X = \tau(t,r,\theta)\partial/\partial{t} + \eta^r(t,r,\theta)\partial/\partial{r} + \eta^\theta(t,r,\theta)\partial/\partial{\theta}$
under which the system \eqref{eom} is infinitesimally invariant, 
\be\label{symmcond}
\pr^{(2)}\X (\ddot r - f^r)|_\Esp = 0,
\qquad
\pr^{(2)}\X (\ddot\theta - f^\theta)|_\Esp = 0 .
\ee
Here $\pr^{(2)}\X$ is the second prolongation of $\X$, 
acting on the coordinate space $(t,r,\theta,\dot r,\dot\theta,\ddot r,\ddot\theta)$. 
The prolongation formula is somewhat complicated. 
It can be avoided by working with the generator $\X$ in evolutionary form
\be\label{pointsymm}
\hat\X = P^r \partial/\partial{r} + P^\theta \partial/\partial{\theta},
\qquad
P^r = \eta^r -\tau \dot r,
\quad
P^\theta = \eta^\theta-\tau \dot\theta,
\ee
acting on solutions $(r(t),\theta(t))$. 
Then the invariance condition \eqref{symmcond} becomes simply
\be
\begin{aligned}
0 & =\pr^{(1)}\hat\X(\ddot r - f^r)|_\Esp = \big( \ddot P^r - P^r \partial_r f^r - P^\theta \partial_\theta f^r - \dot P^r \partial_{\dot r} f^r  - \dot P^\theta \partial_{\dot\theta} f^r \big)\big|_\Esp,
\\
0 & =\pr^{(1)}\hat\X(\ddot\theta -f^\theta)|_\Esp = \big( \ddot P^\theta - P^r \partial_r f^\theta - P^\theta \partial_\theta f^\theta - \dot P^r \partial_{\dot r} f^\theta  - \dot P^\theta \partial_{\dot\theta} f^\theta \big)\big|_\Esp.
\end{aligned}
\ee
This pair of determining equations splits with respect to $\dot r$, $\dot \theta$, 
and thereby yields an overdetermined linear system of equations for $\tau$, $\eta^r$, $\eta^\theta$. 
After simplification, the linear system reduces to 
$\partial_t\tau=\partial_r\tau=\partial_\theta\tau=0$, $\eta^r=0$,
$\partial_t\eta^\theta=\partial_r\eta^\theta=\partial_\theta\eta^\theta=0$,
whose solution is given by $\tau=C_1$, $\eta^r=0$, $\eta^\theta=C_2$,
where $C_1,C_2$ are constants. 
Hence, we obtain the two point symmetries \eqref{X-L-E},
consisting of rotations ($C_1=0$, $C_2=-1$) and time-translations ($C_1=1$, $C_2=0$). 
Both of these point symmetries are variational. 
In particular, in evolutionary form \eqref{hatX-L-E}, 
their action on the Lagrangian is a total time derivative given by 
\be\label{XactionLagr}
\pr^{(1)}\hat\X_{L}(\L) = \dot R =0,
\qquad
\pr^{(1)}\hat\X_{E}(\L) = \dot R = -\dot\L .
\ee

\emph{Step 2}:
Use Noether's theorem to obtain first integrals from the variational point symmetries. 

The action \eqref{XactionLagr} of the variational point symmetries \eqref{hatX-L-E} 
on $\L$ gives $R=0$ and $R=-\L$ respectively. 
Then, through the Noether correspondence \eqref{I}, 
this yields the first integrals $I=L$ and $I=E$, 
which are given by the angular momentum \eqref{L} and the energy \eqref{E}.

\emph{Step 3}:
Re-write the Hamiltonian system \eqref{eom} in first-order form using the previous first integrals. 

First, expressions \eqref{L} and \eqref{E} for the first integrals directly yield 
\begin{equation}\label{1stordsys1}
\dot r = \frac{\sgn(\dot r)\sqrt{2E(\lambda r^2 +1) - L^2/r^2 - \omega^2r^2}}{\lambda r^2 +1} = F^r, 
\qquad
\dot \theta = \frac{L}{r^2(\lambda r^2 +1)} = F^\theta.
\end{equation}
Next, since the first integrals are time-independent, they satisfy 
\begin{equation}\label{1stordsys2}
\dot L =0,
\qquad
\dot E =0.
\end{equation}
These four equations constitute a first-order form for the Hamiltonian system \eqref{eom}.

\emph{Step 4}: 
Find all point symmetries of the first-order system such that
every joint invariant of the variational point symmetries is preserved. 

An invariant of a point symmetry $\X$ is a function $\chi$ of $t,r,\theta$ that is annihilated by the symmetry generator, $\X(\chi)=0$. 
The joint invariants of the two variational point symmetries \eqref{X-L-E} 
clearly consist of only $\chi=r$. 
To preserve this invariant, 
we search for point symmetries on $(t,r,\theta,E,L)$ of the infinitesimal form 
\be\label{Y}
\Y = \tau(r,L,E)\partial/\partial t + \eta^\theta(r,L,E)\partial/\partial\theta + \eta^L(r,L,E)\partial/\partial L + \eta^E(r,L,E)\partial/\partial E.
\ee
The condition for the first-order system \eqref{1stordsys1}--\eqref{1stordsys2} to be infinitesimally invariant is given by 
\begin{gather*}
0 =\pr^{(1)}\hat\Y(\dot L)|_\Esp 
= F^r\partial_r\eta^L,
\qquad
0 =\pr^{(1)}\hat\Y(\dot E)|_\Esp 
= F^r\partial_r\eta^E,
\\
0 =\pr^{(1)}\hat\Y(\dot \theta - F^\theta)|_\Esp 
= \big( F^r(\partial_r\eta^\theta -F^\theta\partial_r\tau) -\eta^E\partial_E F^\theta - \eta^L\partial_L F^\theta \big), 
\\
0 =\pr^{(1)}\hat\Y(\dot r - F^r)|_\Esp 
= -\big( (F^r)^2\partial_r\tau + \eta^E\partial_E F^r + \eta^L\partial_L F^r  \big),
\end{gather*}
using the symmetry generator in evolutionary form 
$\hat\Y= (\eta^\theta -\tau F^\theta)\partial/\partial{\theta} -\tau F^r\partial/\partial{r} + \eta^L\partial/\partial{L} + \eta^E\partial/\partial{E}$.
These determining equations yield the linear system 
\begin{gather}
\partial_r\eta^L=0, 
\quad
\partial_r\eta^E=0,
\\
\partial_r\tau = -(\eta^E\partial_E F^r + \eta^L\partial_L F^r)/(F^r)^2,
\quad
\partial_r\eta^\theta = -(\eta^E\partial_E(F^r F^\theta) +\eta^L\partial_L(F^r F^\theta))/(F^r)^2,
\end{gather}
which can be straightforwardly solved. 
Up to arbitrary functions of $L$ and $E$, 
the general solution is given by 
$\tau = C_1\partial_E\Theta + C_2\partial_E T +C_3$, 
$\eta^\theta = -C_1\partial_L\Theta - C_2\partial_LT -C_4$, 
$\eta^L = C_1$, $\eta^E=C_2$,
where $C_1,C_2,C_3,C_4$ are constants. 
Hence we obtain four point symmetries
\begin{gather}
\Y_L =  -\partial/\partial{\theta} = \X_L,
\qquad
\Y_E = \partial/\partial{t} = \X_E,
\label{Y-L-E}
\\
\Y_\Theta  =\partial_E\Theta\partial/\partial{t} -\partial_L\Theta \partial/\partial{\theta} + \partial/\partial{L},
\qquad
\Y_T  =-\partial_E T\partial/\partial{t} +\partial_L T \partial/\partial{\theta} - \partial/\partial{E}.
\label{Y-Theta-T}
\end{gather}

\emph{Step 5}:
Convert the additional point symmetries into dynamical symmetries of the Hamiltonian system \eqref{eom}. 

The two symmetries $\Y_L$ and $\Y_E$ are clearly inherited from the 
rotation and time-translation symmetries \eqref{X-L-E} of the Hamiltonian system \eqref{eom}. 
Since these symmetries are the only point symmetries admitted by this system, 
the two additional symmetries $\Y_\Theta$ and $\Y_T$ must therefore yield 
dynamical symmetries when they are transformed to act on solutions $(r(t),\theta(t))$.
Their action is obtained simply by first expressing the symmetries in evolutionary form 
and then projecting the generators onto the coordinate space $(r,\theta)$.
This yields 
$\hat\Y_\Theta = -(\partial_L \Theta +\dot\theta\partial_E \Theta)\partial/\partial{\theta} -\dot r\partial_E\Theta\partial/\partial{r} =\hat\X_\Theta$
and 
$\hat\Y_T = (\partial_L T+\dot\theta\partial_E T )\partial/\partial{\theta} +\dot r\partial_E T\partial/\partial{r} =-\hat\X_T$,
which are the two dynamical symmetries \eqref{hatX-Theta} and \eqref{hatX-T} 
obtained previously.

\emph{Step 6}:
Apply Noether's theorem to obtain first integrals by using the dynamical symmetries that are variational. 

There are two methods to verify a priori that the two dynamical symmetries $\hat\Y_\Theta$ and $\hat\Y_T$ are variational, and to derive the corresponding first integrals. 
A direct method \cite{BA-book,Olver-book}
consists of showing that the action of the symmetries on the Lagrangian is a total time derivative, $\pr^{(1)}\hat\Y(\L)=\dot R$. 
This also yields $R$ so that the first integral \eqref{I} can be obtained. 
However, it is somewhat complicated to find $R$ explicitly. 
An alternative method which by-passes this complication 
uses only the dynamical symmetries themselves \cite{BA-book,AncBlu98}. 
First, a dynamical symmetry 
$\hat\Y = P^r\partial/\partial{r} + P^\theta\partial/\partial{\theta}$ is variational iff 
the pair $(Q^r,Q^\theta)$ given by the Noether correspondence \eqref{QfromP} is a multiplier, 
satisfying 
\be\label{Q-deteqns}
\frac{\delta}{\delta r}\Big((\ddot r -f^r)Q^r+(\ddot \theta -f^\theta)Q^\theta\Big)=0,
\quad
\frac{\delta}{\delta \theta}\Big((\ddot r -f^r)Q^r+(\ddot \theta -f^\theta)Q^\theta\Big)=0.
\ee
Then, the resulting first integral \eqref{I} can be obtained by a line integral formula
\be\label{I-lineintegral}
I = \int_{\mathcal C} \Big( (1+\lambda r^2)^{-1}\dot P^r\,dr  + (r^2(1+\lambda r^2))^{-1}\dot P^\theta\,d\theta -(1+\lambda r^2)P^r\,d\dot r -r^2(1+\lambda r^2)P^\theta\,d\dot \theta \Big)\big|_\Esp
\ee
where $\mathcal C$ is any curve from $(t_0,r_0,\theta_0,\dot r_0,\dot\theta_0)$
to $(t,r,\theta,\dot r,\dot\theta)$. 
It is straightforward to verify that the multiplier equations \eqref{Q-deteqns} hold 
for both of the dynamical symmetries $\hat\Y_\Theta$ and $\hat\Y_T$,
using the pairs \eqref{Q-Theta} and \eqref{Q-T}.  
The corresponding first integrals from formula \eqref{I-lineintegral} 
(up to an additive constant) are $I=\Theta$ and $I=T$, 
given by expressions \eqref{Theta} and \eqref{T}, respectively. 

Altogether, the two first integrals arising from the dynamical symmetries, 
plus the two first integrals arising from the point symmetries, 
comprise the complete set of four functionally-independent first integrals
\eqref{L}--\eqref{T} for the Hamiltonian system \eqref{eom}.

\subsection{Variational symmetry algebra}

The commutator structure of the four variational symmetries \eqref{hatX-L-E}--\eqref{hatX-T} 
can be derived in several different ways. 
A direct method is to express the commutators in terms of the action of the prolonged symmetries on the components of the symmetry generators. 
However, the prolongations can be avoided by instead using the representation of the symmetries 
in the form \eqref{Y-L-E}--\eqref{Y-Theta-T} 
given by point symmetries acting on the coordinate space $(t,r,\theta,E,L)$. 
The commutators are then simple to compute, 
and we find that every commutator vanishes. 
Hence the four point symmetries \eqref{Y-L-E}--\eqref{Y-Theta-T} 
comprise an abelian algebra. 
The same algebra structure then holds for the four variational symmetries \eqref{hatX-L-E}--\eqref{hatX-T}. 

We remark that this result provides an alternative way to obtain the corresponding first integrals,
by utilizing the canonical coordinates of the four symmetries \eqref{Y-L-E}--\eqref{Y-Theta-T}
as follows.  
The canonical form of a symmetry \eqref{Y} is given by 
$\Y = \partial/\partial{\zeta}$ 
where $\zeta(t,r,\theta,L,E)$ is a canonical coordinate satisfying 
$\Y(\zeta)=\tau\partial_t\zeta + \eta^\theta\partial_\theta\zeta + \eta^L\partial_L\zeta + \eta^E\partial_E\zeta=1$.
Since the four symmetries \eqref{Y-L-E}--\eqref{Y-Theta-T} are mutually commuting, 
there exists a point transformation from $(t,r,\theta,E,L)$ to the coordinate space 
$(r,\zeta^L,\zeta^E,\zeta^\Theta,\zeta^T)$ 
consisting of the four canonical coordinates and the joint invariant of all four symmetries. 
It is straightforward to see that $\zeta^L = -\Theta$, $\zeta^E = T$, $\zeta^\Theta=L$, $\zeta^T=-E$.

%%%%%%%%%%%%%%%%%%%%%%%%%%%%%%%%%%%%%%%%%%%%%%%%

\section{Concluding remarks}

Finally, we stress that the approach presented in this paper is widely applicable to many other Hamiltonian systems 
and will be helpful for unveiling the connections between local and global aspects of their integrability and symmetry properties.
In particular, we plan to study first integrals and local symmetries of 
more general central force systems with Hamiltonians of the form $H=m(r)^{-1}((p^r)^2 + (p^\theta/r)^2)+ U(r)$
having a position-dependent mass, 
as well as generic (for instance, H\'enon-Heiles type) nonlinearly coupled systems of oscillators. 
Indeed, from a more general perspective, 
an important open problem is how to detect global regularity of first integrals.

%%%%%%%%%%%%%%%%%%%%%%%%%%%%%%%%%%%%%%%%%%%%%%%%%%%

\section*{Acknowledgements}

S.C.A. is supported by an NSERC research grant. 
A.B. has been partially supported by Ministerio de Ciencia, Innovaci\'on y Universidades (Spain) under grant MTM2016-79639-P (AEI/FEDER, UE), by Junta de Castilla y Le\'on (Spain) under grant BU229P18. M.L.G. gratefully acknowledges support by Junta de Andaluc\'{i}a (Spain) for the research group grant FQM-201. 

%%%%%%%%%%%%%%%%%%%%%%%%%%%%%%%%%%%%%%%%%%%%%%%%%%%

\end{document}